\documentclass[aps,pra,twocolumn,superscriptaddress,10pt]{revtex4-1}
\usepackage{amssymb, amsmath,color,graphicx,subfigure}
\usepackage{braket,epsfig,epstopdf,color,bm,mathrsfs,pifont}
\usepackage{times}

\begin{document}

\title{Error-mitigated initialization of surface codes with non-Pauli stabilizers}

\author{Zhi-Cheng He}

\affiliation{Key Laboratory of Atomic and Subatomic Structure and Quantum Control (Ministry of Education), Guangdong Basic Research Center of Excellence for Structure and Fundamental Interactions of Matter, and School of Physics, South China Normal University, Guangzhou 510006, China}

\author{Zheng-Yuan Xue} \email{Corresponding author. Email: zyxue83@163.com}
\affiliation{Key Laboratory of Atomic and Subatomic Structure and Quantum Control (Ministry of Education), Guangdong Basic Research Center of Excellence for Structure and Fundamental Interactions of Matter, and School of Physics, South China Normal University, Guangzhou 510006, China}

\affiliation{Guangdong Provincial Key Laboratory of Quantum Engineering and Quantum Materials,  Guangdong-Hong Kong Joint Laboratory of Quantum Matter,   and Frontier Research Institute for Physics,\\ South China Normal University, Guangzhou 510006, China}
\affiliation{Hefei National Laboratory,  Hefei 230088, China}

\date{\today}

\begin{abstract}
Quantum error correction represents a significant milestone in large-scale quantum computing, with the surface code being a prominent strategy due to its high error threshold and experimental feasibility. However, it is challenging to implement non-Clifford logical gates in a fault-tolerant way with low overhead, through the conventional magic state distillation technique. Here, we enhance the performance of the conventional surface code by incorporating non-Pauli stabilizers and introduce two innovative initialization protocols. Our approach enhances the fidelity of the initialization of non-Clifford logical state  by avoiding unprotected operations before the encoding process. This improved fidelity of the initialization of non-Clifford logical states reduces the necessary number of logical qubits for precise state distillation, ultimately decreasing the overall resource overhead. Furthermore, we demonstrate the ability to entangle logical qubits in non-Pauli and Pauli bases via the lattice surgery technique. This integration enables the use of Pauli-based surface codes for computation while non-Pauli codes are employed for auxiliary qubit initialization, thus compatible with the conventional wisdom of logical Clifford operation based on the Pauli-based surface code.
\end{abstract}

\maketitle

\section{Introduction}
Quantum computers hold the promise of outperforming their classical counterparts in some hard tasks that demanding exponential computational speedups. However, the fragility of quantum information necessitates robust error mitigation and correction techniques for practical large-scale quantum computing. Consequently, various quantum error correction codes (QECCs) have been devised to protect quantum information against primary error sources \cite{QEC_Shor, QEC_Gottesman, QEC_Steane, QEC_Calderbank, QEC_Laflamme, QEC_Bennett, QEC_Cleve, QEC_Knill}.
In QECCs, encoding a single logical qubit typically involves many physical qubits, in order to enable the correction of errors in these physical qubits. Among QECCs, stabilizer codes stand out because of their ability to map various quantum errors to  Pauli errors, thereby simplifying the error detection and correction processes.
The surface code, a prominent stabilizer-based QECC, uses topology-based methods to store information on a 2D lattice structure through local interactions between neighboring qubits. This approach, characterized by its experimental feasibility, has attracted significant attention in various practical implementations, making it a leading choice for the realization of QECC \cite{Surface_code, Surface_code_Kitaev, Surface_code_Raussendorf, Surface_code_high_threshold, Surface_code_error_rate_1, Surface_code_google, Surface_code_real_noise, Surface_code_repeated, Superconducting_to_surface_code, Surface_code_classical_processing}.

The quantum error correction in the surface code is devised to significantly extend the storage of quantum information beyond its physical coherence time, and thus enabling the execution of  quantum circuits composed of Clifford logical-qubit gates in a fault-tolerantly way \cite{braiding_2006, braiding_2007, Lattice_surgery_Horsman, Lattice_surgery_twist, Lattice_surgery_game, Lattice_surgery_experiment,
Topological_Order_with_a_Twist, Poking_holes_and_cutting_corners, surface_code_with_a_twist}. However,  the implementation of non-Clifford gates in a fault-tolerant and low-overhead way, which is essential to demonstrate the advantage of quantum computers, remains challenging \cite{Universal_1, Universal_2}. The conventional  approach for obtaining non-Clifford gates involves fault-tolerantly initializing an auxiliary logical qubit into a non-Clifford state, which requires the use of the magic state distillation technique
\cite{Magic_state_distillation_low_overhead,Magic_state_distillation_not_costly}. But, this process is resource-intensive and dominates the overall resource cost of the surface code-based quantum computation. Therefore, to mitigate these costs and facilitate fault-tolerant quantum computation, it is crucial to develop initialization protocols for logical qubits with a error rate as low as possible. By doing so, one can minimize the required number of noisy logical qubits, thereby decreasing the qubit resource overhead associated with the magic state distillation process. Recently, to approach lower logical-state error rate before distillation, several protocols for initializing non-Clifford states have been proposed \cite{break-even, Analog_Rotations, Lis_paper, Preparation_ZKD, post_selection, Transversal_Injection}. However, most of the protocols rely on single-qubit operations to inject the target quantum information into the logical state, thereby rendering such errors undetectable by the QECC.

Here, we propose a novel strategy for encoding arbitrary logical qubit states for quantum computation based on the surface code that bypasses the conventional noisy injection step, using non-Pauli stabilizers \cite{Non-Pauli_topological_stabilizer_codes, XP_stabilizer_Formalism}, thus makes it particularly advantageous for the Pauli stabilizer based QECC. The fundamental principle of this approach is that the ground state can be represented as a superposition state on a prescribed non-Pauli basis. Consequently, when processing on this non-Pauli basis, the corresponding logical information is encoded into the lattice of physical qubits. Our protocols can tolerate single-qubit errors, thereby enhancing initialization fidelity and reducing resource overhead for the magic state distillation. In particular, a non-Pauli stabilizer based logical qubit can be entangled with that of the Pauli stabilizer based qubit via the lattice surgery technique. This ensures the use of Pauli stabilizer to store and entangle quantum information, while the non-Pauli logical qubits facilitate non-Clifford operations.

\begin{figure*}[tb]
\centering \includegraphics[width=1.0\textwidth]{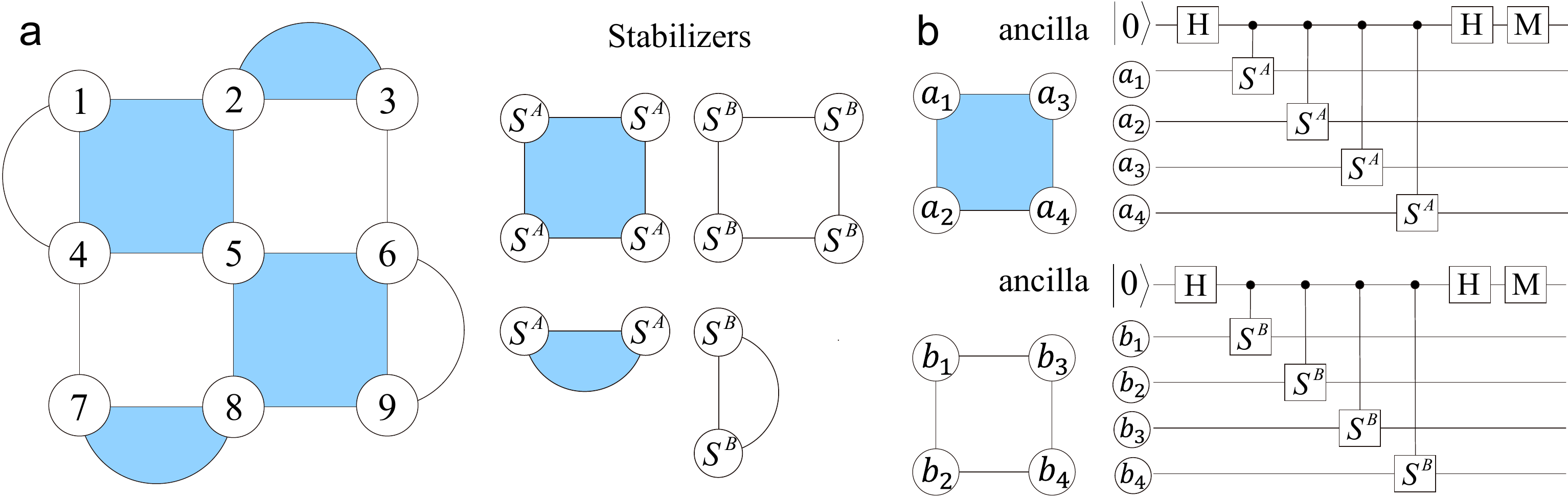}
\caption{ Illustration of our scheme. a) Example for a distance 3 rotated surface code with non-Pauli stabilizers. The blue (white) tiles represent the $S^{A}$ ($S^{B}$) -type stabilizers. The logical operators respect to this layout can be $X_{L}=S^{A}_{1}S^{A}_{4}S^{A}_{7}$ and $Z_{L}=S^{B}_{1}S^{B}_{2}S^{B}_{3}$.  b) Quantum circuit to realize the stabilizer measurement. The ancillary qubits lived in the center of tiles have been omitted.}
\label{Lattice_and_circuit}
\end{figure*}

\section{The surface code with non-Pauli operators}

In this section, we first introduce the definition of the Non-Pauli stabilizer, from which we then construct a new surface code that can be capable for fault-tolerant quantum computation. Finally, we present the error model of the proposed surface code with Non-Pauli operators.

\subsection{Non-Pauli stabilizers}
In quantum computation, stabilizer codes represent a fundamental class of quantum codes used to correct quantum errors. Within the stabilizer code framework, the logical subspace is characterized as the simultaneous eigenspace of a set of mutually commuting product of Pauli operators. These products of Pauli operators, referred to as stabilizers, collectively constitute an Abelian stabilizer group. Several reasons underlie the choice of Pauli operators for generating stabilizers. Firstly, Pauli operators possess the property $X^{2}=Y^{2}=Z^{2}=I$, which facilitates  closure under multiplication, a key requirement for forming a stabilizer group. Secondly, Pauli operators acting on the same qubit anti-commute, i.e., $AB=-BA$, for distinct Pauli operators $A$ and $B$. These anticommuting relations enable the design of commuting stabilizers even when different operators are applied to the same qubit. For example, the seven-qubit Steane code comprises six stabilizers, two of which can be expressed as $XXXXIII$ and $ZZZZIII$ on seven qubits. It is noteworthy that these two stabilizers commute due to $[X_{i}X_{j}, Z_{i}Z_{j}]=0$ for qubits $i,j$, a consequence of the anti-communication relation of $XZ=-ZX$ for a same qubit.

While Pauli operators are commonly used due to their convenience in expression and analysis, other two-level operators can also generate the stabilizer group. Here, we introduce two such operators, denoted as $S^{A}$ and $S^{B}$, where $S^{n}$, with $n \in \{ A, B\}$, represents an arbitrary two-level operator distinct from the Pauli type. In stabilizing a computational subspace, we observe that all Pauli operators acting on a single qubit can be substituted with non-Pauli operators $S^{A}$ and $S^{B}$, constituting new commutative stabilizers, provided that the following conditions are met: $(S^{A})^2=(S^{B})^2=I$ and $S^{A}_{i}S^{B}_{j}=-S^{B}_{i}S^{A}_{j}$ for any pair of qubits ${i,j}$. The logical gate operations and the measurement circuits still work normally when the corresponding operators are switched from Pauli to non-Pauli types. For example, in the case of the six stabilizers of the Steane code, we can substitute the Pauli operators on the first qubit with non-Pauli operators, $X\rightarrow{}S^{A}$ and $Z\rightarrow{}S^{B}$. This substitution results in a modification of all six stabilizers, such as the first two stabilizers changing to $S^{A}XXXIII$ and $S^{B}ZZZIII$. These transformed stabilizers remain commutative after substitution due to the commutativity $[S^{A}_{i}X_{j},S^{B}_{i}Z_{j}]=0$ for qubits $i,j$.

\subsection{The surface code with non-Pauli operators}

The lattice configuration for the proposed surface code based on non-Pauli operators is schematically shown in Fig. \ref{Lattice_and_circuit}a, where a distance 3 case is presented for illustration purposes, and the nearby stabilizers overlap on two data qubits and two different types of stabilizer are defined. The logical subspace is an common eigenspace that is shared by all stabilizers, thus requiring the overlapping data qubits of different stabilizers to be commuted. A conventional approach is to use operators that satisfy the anticommute relation for all data qubits, such as the Pauli operators $X$ and $Z$ for the surface code scheme. Here, we employ arbitrary two-level operators $S^{A}$ and $S^{B}$ for a qubit, satisfying anticommute relation $S^{A} S^{B} =-S^{B} S^{A}$, to construct the needed two types of stabilizers, i.e., $S^A$-type and $S^B$-type. In general, $S^{A}$ and $S^{B}$ are not Pauli operators, and thus the stabilizers composed of them are termed non-Pauli stabilizers, which can be written as $\prod_{i\in\partial f_{A}} S^A_{i}$ and $\prod_{j\in\partial f_{B}} S^B_{j}$, where $f_{A}$ and $f_{B}$ denote the square constructed by the four or two qubits that belongs to the $S^A$-type or $S^B$-type, respectively.

For clarity, we express the matrix form of $S^A$ and $S^B$, using operators transformed form Pauli operators, as
\begin{equation}
S^A=UXU^{\dagger},\quad
S^B=UZU^{\dagger},
\end{equation}
with $U$ being a transformation operator as
\begin{equation}
\label{U}
U=\exp\left[{\text{i}\frac{\gamma}{2}
\left(
\begin{matrix}
\cos{\theta} & \sin{\theta}e^{-i\varphi} \\
\sin{\theta}e^{i\varphi} & -\cos{\theta}
\end{matrix}
\right)} \right],
\end{equation}
where $\gamma$, $\theta$ and $\varphi$ describe a full $SU(2)$ transformation.
Note that these new stabilizers still satisfy the anticommutation condition since $[(UXU^{\dagger})_{i}(UXU^{\dagger})_{j}, (UZU^{\dagger})_{i}(UZU^{\dagger})_{j}]=0$.
Consequently, with respect to this transformation,  the eigenstates of them are
\begin{equation}
\ket{\pm}_{A}=U\ket{\pm}_{X}, \quad
\ket{\pm}_{B}=U\ket{\pm}_{Z},
\end{equation}
where $\pm$ and $\ket{\pm}$ denote positive or negative eigenvalues and eigenstates  of operators $\{S^{A},S^{B},X,Z\}$.

To facilitate the measurement of non-Pauli stabilizers, we employ the Hadamard test circuit \cite{Nielsen1}, as illustrated in Fig. \ref{Lattice_and_circuit}b.
When the circuit ends, an auxiliary qubit is measured. After that, the state of four data qubits will collapse to the positive value eigenstate of the stabilizer in the circuit, if the ancillary qubit is measured in the ground state. Otherwise, if the ancillary qubit is measured in the excited state, the state collapses to a negative value eigenstate of corresponding stabilizer.
Due to the Hadamard test circuit enables measuring all entangled unitary two-level operators, it can be used to measure all possible $S^n$-type stabilizers in our protocol.
We note that, in the measure of $S^n$-type stabilizers, implementation of the corresponding control-$S^n$ gate is required, which can be directly obtained through Hamiltonian designs. For Rydberg atoms, the strong Rydberg-Rydberg interaction can implement a control-$U$ gate for arbitrary $U$. For the implementation of control-$U$ gates, the Hamiltonian probably are effected by environment or control noises. These noises are usually described as Pauli matrices \cite{Saffman, LiangYan}. Here, we assume that the error rate is same for all control-$U$ gates, since different control-$U$ gates will achieve almost identical average fidelities when against Pauli noises.

\subsection{The error model}

Errors in quantum computation typically manifest as arbitrary two-level operators in the computational subspace of physical qubits, arising from the inevitably imperfections of experimental control or from decoherence. One of the most powerful features of stabilizer codes is their ability to project arbitrary errors onto Pauli errors through stabilizer measurements. This property enables the consideration of Pauli errors alone in general quantum error analysis and simulations, thus leads to the twirled channel,
\begin{equation}
\label{Pauli error}
\mathcal{N}'(\rho)=(1-p')I\rho I+p_x X\rho X+p_z Z\rho Z+p_y Y\rho Y,
\end{equation}
where $p'=p_x+p_y+p_z$ is the overall error rate for a physical qubit,  $p_{i}$ with $i\in\{x, y, z\}$ is the error rate of the qubit in $i$ direction, and the value of $p_{i}$ depend on the different type of quantum channels. The measurement of a stabilizer can be viewed as an operation that projects $\rho$ to a new density matrix which undergoes an identity gate with a probability of $(1-p')$, or one of Pauli errors $i$ with a probability of $p_{i}$.
The ability to establish Eq. (\ref{Pauli error}) is due to the Pauli basis is a completely-positive, trace-preserving complete linear map for  two-level quantum systems.

Similarly, for the case of non-Pauli stabilizer scheme, the error components can be conveniently decomposed  in the corresponding non-Pauli basis. That is,
\begin{equation}
\label{non-Pauli error}
\mathcal{N}(\rho)=(1-p_s)I\rho I+p_{a} S^A\rho S^A+p_{b} S^B\rho S^B+p_{c} S^C\rho S^C,
\end{equation}
where $p_s=p_a+p_b+p_c$ is the single-physical-qubit error rate and $S^{C}=iS^{A}S^{B}$. Note that, this change of basis does not affect the threshold value for the surface code.
For the error detection process, the non-Pauli stabilizer scheme  also provides the same capabilities as the Pauli scheme. Specifically,  if a projected error $S^{A}$ occurs on a data qubit, the eigenstate of corresponding $S^{B}$-type stabilizer changes to another eigenstate with a different eigenvalue. Thus, the occurrence of an $S^{A}(S^{B})$ error would be detected by $S^{B}(S^{A})$-type stabilizers. Furthermore, if the eigenvalues of both the $S^{A}$-type and $S^{B}$-type stabilizers change, it indicates that both $S^{B}$ and $S^{A}$ error occurred on the corresponding qubit, i.e., the $S^{C}$ error is detected, as
$S^{C}=iS^{A}S^{B}$.

\begin{figure}[t]
\centering
\includegraphics[width=\columnwidth]{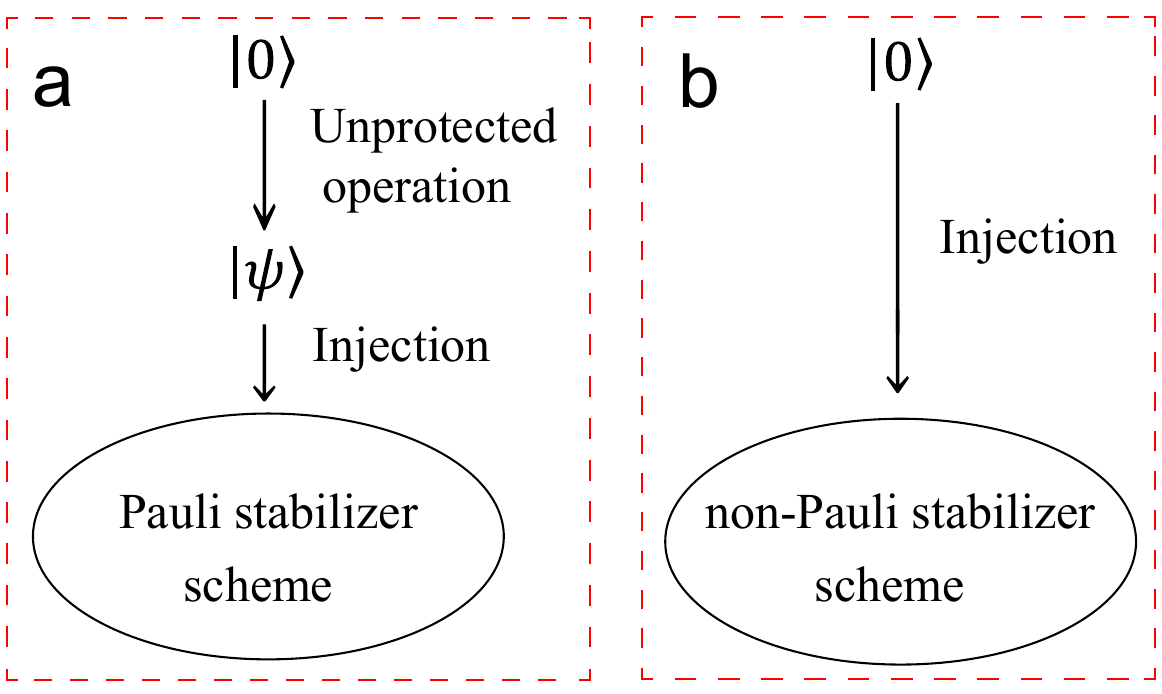}
\caption{Different initialization scheme for Pauli and non-Pauli stabilizer schemes. a) The initialization scheme for the Pauli stabilizer code. Here, it requires vulnerable single-qubit gates to obtain physical non-Clifford states and then inject the information to the code space. The error occurred in vulnerable operation can not be detected. b) The initialization scheme for the non-Pauli stabilizer scheme. The ground state can be injected to non-Pauli code space directly since it is a superposition state in non-Pauli basis with corresponding superposition coefficients.}
\label{Injection}
\end{figure}

\section{Initialization of logical qubits}
In this section, we first proposed two initialization schemes for logical qubits under the proposed surface code, and then analysis their performance.

\subsection{The protocol}
Universal quantum computation necessitates a fault-tolerant protocol to initialize quantum states. Although initializing logical Clifford states is relatively straightforward for most QECCs, initializing logical non-Clifford  states is still challenging due to error tolerance constraints. Given that qubits naturally decay to the ground state, initialization will be preferred to begin with all qubits in this state, which is an eigenstate of the Pauli $Z$ operator. Consequently, the absence of non-Clifford superposition information in a Pauli-based architecture is inherent. To create a logical magic state, as shown in Fig. \ref{Injection}a, it is essential to apply single-qubit gates to introduce non-Clifford information and subsequently expand it across the array. However, these gates are inherently non-fault-tolerant, as errors before the code execution can not be detected and corrected by the code. As shown in Fig. \ref{Injection}b, our research breakthrough involves eliminating single-qubit gates by incorporating non-Pauli stabilizers. This insight stems from the understanding that the ground state is not only a $Z$-eigenstate but also a superposition in a non-Pauli basis. Thus, by designing stabilizers and code spaces using non-Pauli operators, we can achieve arbitrary superposition coefficients in the non-Pauli basis without altering their ground states. Two-qubit operations involving non-Pauli qubits are adapted accordingly. This approach effectively encodes non-Clifford quantum information into the code subspace of qubits without single-qubit control, thereby enhancing fault tolerance by removing unprotected single-qubit operations. In the following, we will describe two distinct methods for initializing arbitrary logical states within a non-Pauli stabilizer framework, differentiating them based on whether they involve pre-measurement on chains within the logical qubit.

The first initial protocol inspired by post-selection-based initialization and a non-Pauli stabilizer approach \cite{post_selection}. To demonstrate this, consider a qubit array for a distance 3 rotated surface code, with all physical qubits in their ground state. We proceed by selecting a specific $X_L$ logical chain, consisting of qubits 1, 4, and 7, to extract the logical information. This information is then propagated from the chain to the entire surface. Initially, the qubits 1, 4, and 7, being in their ground states, can be written as
$\ket{\psi}_{1}\ket{\psi}_{4}\ket{\psi}_{7}=\ket{g}^{\otimes3}$,
where $\ket{\psi}_{i}$ describe the state of the $i$th physical qubit, as shown in Fig. \ref{Lattice_and_circuit}. As the eigenstates of a non-Pauli operator are also complete for representing any two-level state, we choose the basis of $S^{B}$, instead of Pauli $Z$ basis, to represent these ground states,
\begin{equation}
\label{147_before_L}
\ket{\psi}_{1}\ket{\psi}_{4}\ket{\psi}_{7}=(\cos{\theta}\ket{+}^{B}+\sin{\theta}e^{i\varphi}\ket{-}^{B})^{\otimes3},
\end{equation}
where the superposition coefficients  $\theta$ and $\varphi$ are determined by the logical quantum information to be encoded. These coefficients depend on the unitary transformation operator $T$ in Eq. (\ref{U}). And, once  $T$ is determined, the forms of stabilizer operators $S^{A}$, $S^{B}$ and the  coefficients are fixed, which means the non-Clifford information is effectively injected. Next, we  implement the measurement of  temporary stabilizers $S^{B}_{i}S^{B}_{j}$, which covering all nearby pairs of qubits $i,j$ along the target chain, to exact the desired logical information. When the measurement result  become $S^{B}_{1}S^{B}_{4}=1$ and $S^{B}_{4}S^{B}_{7}=1$, the logical chain will collapse to the desired state of
\begin{equation}
\label{147L}
\ket{\psi}_{147}=\frac{1}{\sqrt{N}} \left(\cos^3{\theta}\ket{+++}^{B}_{147}+ \sin^3{\theta}e^{i3\varphi}\ket{---}^{B}_{147}\right),
\end{equation}
where $N$ is the new normalization factor. The process requires post-selection to ensure that the temporary stabilizers' measurement outcomes confirm the system's collapse into the desired entangled state. If the measurements do not match the expectations, the process will be restarted.

The target information in the form of Eq. (\ref{147L}) is then expanded from the chain to the entire surface by measuring all the stabilizers. Before measurements, all other physical qubits beyond the logical chain are driven into the  $\ket{+}_{A/B}$ eigenstate of the operator $S^{A/B}$. Note that errors in this step can be tolerated, conditioned on some special stabilizers reporting the measurement value as 1. If a unexpected measurement value is reported, the process should be restart; the details analysis is demonstrated in the next subsection. Finally, the non-Clifford logical state is ultimately prepared by measuring all the stabilizers, given by
\begin{equation}
\ket{\psi}_{L}=\frac{1}{\sqrt{N}} \left(\cos^3{\theta}\ket{+}^{B}_{L}+ \sin^3{\theta}e^{i3\varphi}\ket{-}^{B}_{L}\right).
\end{equation}
In contrast, employing the general state injection protocol within a Pauli stabilizer framework would require single-qubit gates to achieve the state in Eq. (\ref{147_before_L}) and two-qubit gates for Eq. (\ref{147L}) on the basis of Pauli $Z$. Any error at this stage would result in an unintended logical state.

The second protocol, which integrates the non-Pauli stabilizer scheme and the transversal injection method \cite{Transversal_Injection}, also begins with the qubit lattice all in their ground states, denoted as $\ket{g}^{\otimes n}$. Similar to the previous method, this protocol employs a non-Pauli basis representation for qubits initially in the ground state,
\begin{equation}
\ket{g}^{\otimes n}=\left(\cos{\theta}\ket{+}^{B}+\sin{\theta}e^{i\varphi}\ket{-}^{B}\right)^{\otimes n}.
\end{equation}
Then, all the non-Pauli stabilizers are measured one by one and obtain a logical state. The measurement of each non-Pauli stabilizer generates a sequence, known as the stabilizer trajectory. For instance, in a d=3 rotated surface code, if all 8 measurements for $S^{A}$ and $S^{B}$ stabilizers yield zero, the trajectory would be $\{00000000\}$. After measuring the stabilizers, the resulting logical state, which depends on both the stabilizer trajectory and the specific form of non-Pauli stabilizers, is derived. The corresponding superposition coefficient of the logical state can be obtained by simulating the process on a classical computer beforehand. It should be declared that although all of measurement are probabilistic, the resultant encoded state is still heralded and computable when the code distance increased. This is because the different stabilizer trajectories usually indicate same logical states \cite{Transversal_Injection}. In contrast, the  Pauli-stabilizer  method  needs physical single-qubit with the number $d^{2}$, which could cause failure of initialization.

\begin{figure}[t]
\centering    \includegraphics[width=\columnwidth]{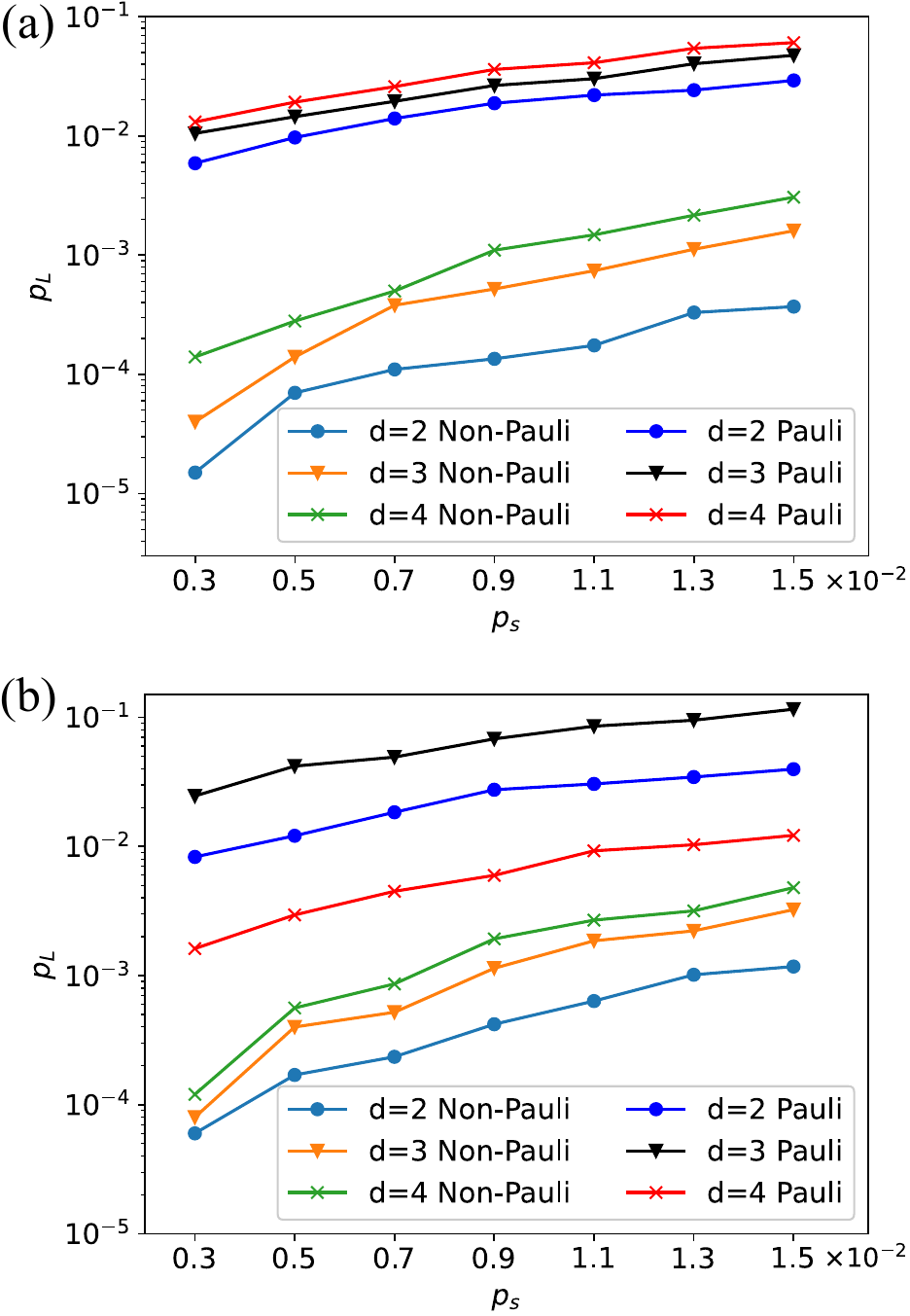}
\caption{Comparison on the performance of different initialization schemes. The performance is evaluated by logical error rate changed with different single-qubit gate error rates in Pauli stabilizer or non-Pauli stabilizer schemes for distance $\{2,3,4\}$ surface codes.  a) Post-selection based initialization of rotated surface code. b) Transversal injection initialization of rotated surface code.}
\label{simulation}
\end{figure}

\subsection{The performance}

In the process of preparing logical states with surface code, there exist three types of physical errors which could lead to the final logical state error, i.e., the two- and single-qubit gate error (denote as $p_t$ and $p_s$), and the reset error ($p_r$) \cite{Lis_paper,Preparation_ZKD}. While the most previous works in initializing logical qubits are vulnerable to single-qubit gate error, in the present non-Pauli based preparation, since the unprotected single-qubit rotation for obtaining the superposition state is no longer needed, and thus makes it advantageous over the Pauli scheme. Here, we focus on the fault tolerance of two protocols respect to the single-physical-qubit error $p_s$ in the initialization process, while neglecting other errors. This is because the Pauli and non-Pauli surface codes have the same performance with respect to other types of error, i.e., the two-physical-qubit errors and reset errors.
To illustrate the fault tolerance of our protocol, we performed extensive simulations of non-Clifford state initialization within various stabilizer schemes, calculating the logical error rates with different physical error rates, using Qiskit \cite{Qiskit}. In the numerical simulation, we employ the general circuit noise model \cite{Surface_code, Lis_paper, Preparation_ZKD}, where every single-qubit gate is assumed to be performed perfectly, then an error channel operation is performed to simulate practical gate errors. If an error does not occur, the error channel operation implements an identity. And, if an error occurs, an error gate is implemented after the gate operation. For the measurement step, we measure stabilizers twice for detecting the single-qubit error in measurement circuits. If the two  measurement results are different, the prepared logical state will be discard. Additionally, the numerical data presented is collected by reusing only one ancillary qubits for the measurement of all stabilizers.
For the first protocol, we chose to initialize the state $\ket{A}_{L}=\frac{1}{\sqrt{2}}(\ket{0}_{L}+e^{i\frac{\pi}{4}}\ket{1}_{L})$. And for second protocol, we firstly excite all the qubits to the state $\ket{A}_{L}$ or keep them in the ground state, respecting to Pauli or non-Pauli stabilizers. We implement the two protocols and finally check the obtained logical state whether it exactly matches the logical state prepared in the absence of errors. Afterward, we counted the number of failed attempts and divided it by the total number of initialization attempts to obtain the logical error rates, which highlight the advantages of our non-Pauli stabilizer schemes. As smaller distance surface codes employ fewer single-qubit gates and thus have lower error rates, the performance with smaller code distances are simulated with more shots to obtain sufficiently precision. Specifically, with the promise of ${10}^{-4}$ precision, we simulate the initialization of a distance 4, 3 and 2 surface code for ${10}^4$,  $5\times{10}^4$ and $2\times{10}^5$ times, respectively. And, both Pauli and non-Pauli schemes are set to have the same number of simulation shots for different code distances.
The results show that all cases in our protocol outperform conventional Pauli schemes, as our scheme tolerant the single-physical-qubit error and thus obtains the lower logical-state error rates, shown in Fig. \ref{simulation}.

\begin{figure*}[t]
	\centering
 \includegraphics[width=\textwidth]{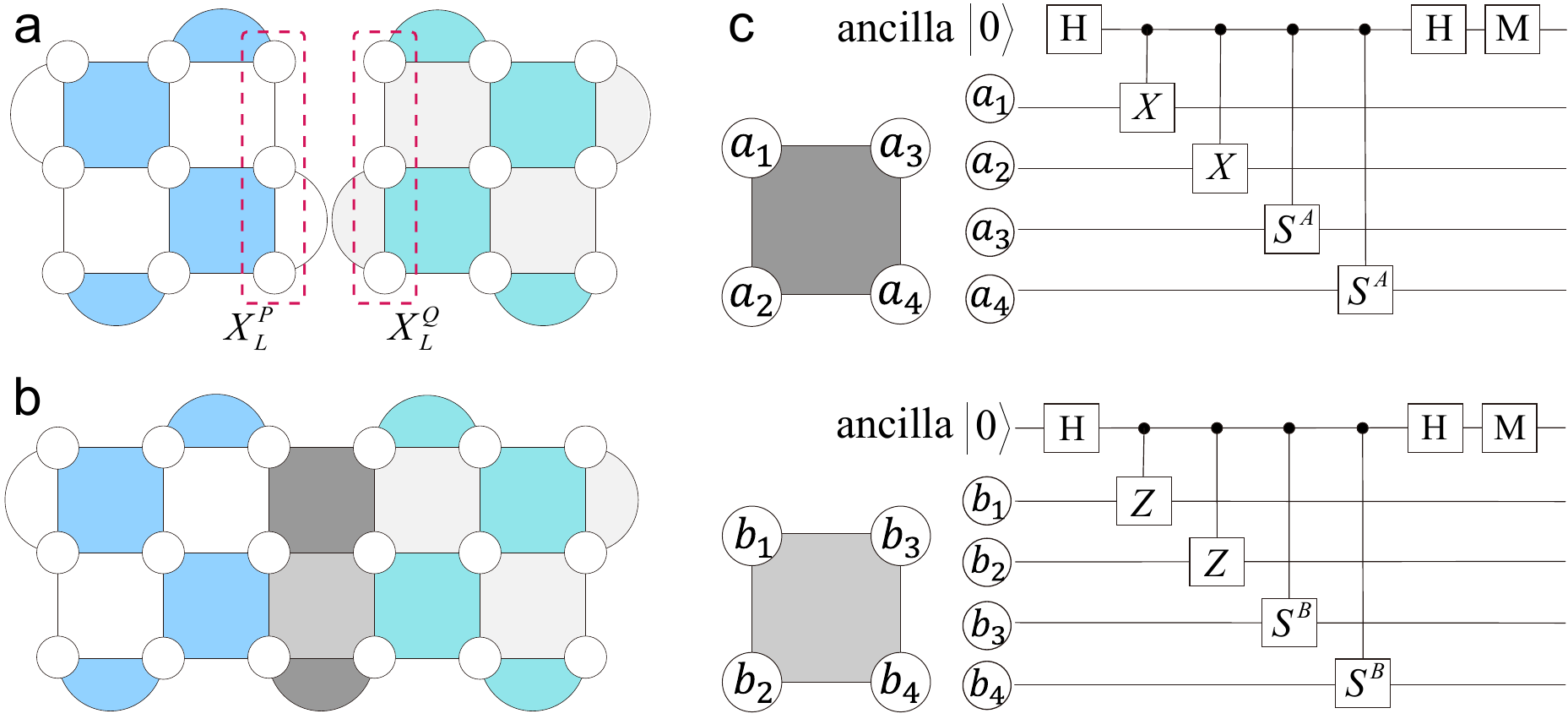}
\caption{ The entangling process for logical qubits based on non-Pauli and Pauli stabilizers via lattice surgery.  a) The layout of non-Pauli stabilizer surface code (left) and Pauli stabilizer surface code (right) before lattice merging. b) The two merged logical qubits in surface codes. The measurement results of the middle deep black stabilizers would be used to obtain the value of logical operators $X^{P}_{L}\otimes X^{Q}_{L}$ for the left and right logical qubits. c) Quantum circuit to realize stabilizer measurement of added middle stabilizers.}
    \label{surgery}
\end{figure*}

Next, we discuss the fault tolerance for the single-qubit errors during the initialization, i.e., the errors that occurred when driving the qubits to the $\ket{+}_{A/B}$ eigenstate, which are relevant to the first post-selection-based protocol. Following the logical chain preparation, qubits not part of the chain are manipulated into the eigenstates of non-Pauli operators, thus preventing unintended collapses that cause the logical state error. Most driving operations exhibit fault tolerance against first-order single-qubit gate errors, as incorrect states of qubits would be projected to the eigenstates of stabilizers after their measurements. This error projection parallels the standard fault-tolerant initialization of trivial logical states $\ket{\pm}_{L}$ or $\ket{1}_{L}$, which transversely drives all qubits into $\ket{\pm}$ or $\ket{1}$ and then measures the stabilizers. However, the measurement process is unable to rectify an incorrect state, which subsequently leads to logical errors, when a stabilizer circuit encounters these two qubits: one within the logical qubit chain, and another outside the chain with errors. Fortunately, the logical error arising from this situation can be mitigated to second-order of physical errors by restarting the preparation whenever the deterministic stabilizer of the first stabilizer cycle yields a incorrect value. Specifically, in the context of stabilizers of type $S^{B}$, a temporary stabilizer $S^{B}_{i}S^{B}_{j}$ and the surface code stabilizer $S^{B}_{i}S^{B}_{j}S^{B}_{k}S^{B}_{l}$ covering it are commutative since they share two overlapping qubits. Assuming the remaining data qubits, qubits $k$ and $l$, have already been projected into the eigenstate of $S^{B}$ operators, their combined state naturally aligns with the eigenstate of the product $S^{B}_{i}S^{B}_{j}\otimes S^{B}_{k}\otimes S^{B}_{l}$. This product state is equivalent to an eigenstate of the $S^{B}$-type stabilizer $S^{B}_{i}S^{B}_{j}S^{B}_{k}S^{B}_{l}$. Therefore, the result of this stabilizer measurement is deterministic in the absence of errors. When the stabilizers report incorrect values, the entire initialization process can be discarded and restarted, thereby achieving tolerance to single-qubit control errors. Secondly, the measurement process also involves a single bit gate in the measurement circuits. These types of error occurred on ancillary qubits, and thus can be detected in the next stabilizer cycle, since that determining the value of a specific measurement requires multiple surface code cycles.

\section{Lattice surgery}

Lattice surgery is a mature fault-tolerant protocol that only requires the measurement of local stabilizer operators to implement a logical entangling gate \cite{Lattice_surgery_Horsman,Lattice_surgery_game,Lattice_surgery_twist, Lattice_surgery_experiment}. For two logical qubits
$\mathcal{P}$ and $\mathcal{Q}$ with logical operators $X^{\mathcal{P}}_{L}$, $Z^{\mathcal{P}}_{L}$ and $X^{\mathcal{Q}}_{L}$, $Z^{\mathcal{Q}}_{L}$, lattice surgery is introduced as a method to project two logical qubits $\mathcal{P}$ and $\mathcal{Q}$ onto the joint eigenstates of $X^{\mathcal{P}}_{L}\otimes X^{\mathcal{Q}}_{L}$ or $Z^{\mathcal{P}}_{L}\otimes Z^{\mathcal{Q}}_{L}$, when the lattice boundary being rough and smooth, respectively. In Fig. \ref{surgery}a, we illustrate the rough boundary case. To obtain this projecting  operation, as shown in Fig. \ref{surgery}b, two lattices of surface code logical qubits will merge to one by measuring the stabilizers at the composed by the two nearby boundary qubits, as indicated by the three dark shapes in Fig. \ref{surgery}a, and then split it to two separately lattices. The product of these $X$ or $Z$ type stabilizers yields the eigenvalue of the joint logical operator $X^{\mathcal{P}}_{L}\otimes X^{\mathcal{Q}}_{L}$ or $Z^{\mathcal{P}}_{L}\otimes Z^{\mathcal{Q}}_{L}$, depending on whether the merged boundary is rough or smooth. Furthermore, a logical control-NOT gate between the control qubit $\mathcal{P}$ and the target qubit $\mathcal{Q}$ can be realized by applying two suitable joint measurements: a parity measurement of $Z^{\mathcal{P}}_{L}\otimes Z^{\mathcal{R}}_{L}$ between the control qubit $\mathcal{P}$ and an auxiliary logical qubit $\mathcal{R}$ initialized into $\ket{+}_{L}$, followed by a parity measurement of $X^{\mathcal{R}}_{L}\otimes X^{\mathcal{Q}}_{L}$ between the auxiliary logical qubit $\mathcal{R}$ and the target qubit $\mathcal{Q}$. That is, these joint measurements of logical operators can entangle the logical qubits, and thus, to achieve efficient universal quantum computation with the non-Pauli surface codes.

We also note that our non-Pauli surface code can be compatible with the Pauli surface codes by the lattice surgery between Pauli and non-Pauli surface codes. This hybrid lattice surgery can be implemented by replacing generic stabilizers with mixed ones, as shown in Fig. \ref{surgery}c. These mixed stabilizers consist of $X$/$Z$ and $S^A$/$S^B$ operators acting on qubits with the surface code based on  Pauli- and non-Pauli- stabilizers, respectively. Consequently,    stabilizers that emerged two logical qubits $P$ and $Q$ are in the form of $X_{P}X_{P}S^{A}_{Q}S^{A}_{Q}$ or $Z_{P}Z_{P}S^{B}_{Q}S^{B}_{Q}$. The corresponding circuit to perform the measurement of these mixed stabilizers is shown in Fig. \ref{surgery}c.
For data qubits in the Pauli-stabilized code, a controlled-NOT (CNOT) or controlled-$Z$ gate is employed to gauge the $X$ or $Z$ operator, while for non-Pauli-stabilized code qubits, a controlled-$S^A$ or controlled-$S^B$ gate is utilized to assess the $S^A$ or $S^B$ operator, respectively.

We next detail the process of integrating two logical qubits via the lattice surgery, with qubits $\ket{\psi}_{\mathcal{P}}=a_{p}\ket{0}_{\mathcal{P}}+b_{p}\ket{1}_{\mathcal{P}}$ and $\ket{\psi}_{\mathcal{Q}}=a_{Q}\ket{0}_{\mathcal{Q}}+b_{Q}\ket{1}_{\mathcal{Q}}$ in the Pauli and non-Pauli basis, respectively.
As an example for lattice merging in rough boundary, we firstly measure the added new mixed stabilizers, and repeat the measurement till it is reliable. The product of the values for new mix stabilizers thus become the eigenvalue of product logical operator $X^{\mathcal{P}}_{L}\otimes X^{\mathcal{Q}}_{L}$. After the measurement, the two qubit state will reduce to
\begin{equation}
\frac{1}{\sqrt{2}}(\ket{\psi}_{\mathcal{P}}\otimes\ket{\psi}_{\mathcal{Q}}+(-1)^{M}X^{\mathcal{P}}_{L}\ket{\psi}_{\mathcal{P}}\otimes X^{\mathcal{P}}_{Q}\ket{\psi}_{\mathcal{Q}})
\end{equation}
where $M$ is the outcome of the logical measurement $X^{\mathcal{P}}_{L}\otimes X^{\mathcal{Q}}_{L}$. Then, consider a new logical operator $Z_{L}$ is a simple product of $Z^{\mathcal{P}}_{L}$ and $Z^{\mathcal{Q}}_{L}$, we can map the two logical state to the new logical state of
\begin{equation}
\begin{aligned}
\ket{0}_{L}=\frac{1}{\sqrt{2}}(\ket{0}_{\mathcal{P}}\ket{0}_{\mathcal{Q}}+(-1)^{M}\ket{1}_{\mathcal{P}}\ket{1}_{\mathcal{Q}}),\\
\ket{1}_{L}=\frac{1}{\sqrt{2}}(\ket{0}_{\mathcal{P}}\ket{1}_{\mathcal{Q}}+(-1)^{M}\ket{1}_{\mathcal{P}}\ket{0}_{\mathcal{Q}}).
\end{aligned}
\end{equation}
Performing the mapping, we get the final state of the lattice merging, with $X^{\mathcal{P}}_{L}\otimes X^{\mathcal{Q}}_{L}$ boundary, as,
\begin{equation}
\begin{aligned}
\ket{\psi}_{\mathcal{P}}\textcircled{m}\ket{\psi}_{\mathcal{Q}}
&=a_{P}\ket{\psi}_{\mathcal{Q}}+(-1)^{M}b_{P}X^{\mathcal{Q}}_{L}\ket{\psi}_{\mathcal{Q}}\\
&=a_{Q}\ket{\psi}_{\mathcal{P}}+(-1)^{M}b_{Q}X^{\mathcal{P}}_{L}\ket{\psi}_{\mathcal{P}},
\end{aligned}
\end{equation}
where we employ the symbol $\textcircled{m}$ to label lattice merging operation \cite{Lattice_surgery_Horsman}.
Note that the entangling operation is just based on the measurement of joint logical operator $X^{\mathcal{P}}_{L}\otimes X^{\mathcal{Q}}_{L}$. And, this types of measurement would not change the logical state to a disorder one, the merging still work perfectly even the basis of two logical code is different. With this hybrid lattice surgery, we can use Pauli-based surface codes for computation while non-Pauli codes are employed for auxiliary qubit initialization, thus maintaining the conventional wisdom for logical Clifford operation.

\section{Conclusion}
In conclusion, we present a comprehensive non-Pauli stabilizer scheme that facilitates the initialization of non-Clifford logical qubits within surface codes, while eliminating unprotected injecting operations and leading a much higher fidelity. Furthermore, the resource overhead for initializing logical qubits associated with this protocol remains comparable to traditional initialization techniques. Future efforts could thus be directed towards reducing the error rates associated with two-qubit gates or enhancing the detection of errors occurring during the execution of these gates.

\acknowledgements
We thank Dr J. Liu for helpful discussions.
This work was supported by the National Natural Science Foundation of China (Grant No. 12275090), the Guangdong Provincial Quantum Science Strategic Initiative (Grant No. GDZX2203001), and the Innovation Program for Quantum Science and Technology (Grant No. 2021ZD0302303).

\end{document}